\documentclass[usegraphicx, useAMS, usenatbib, letterpaper]{mn2e}

\usepackage{graphicx}
\usepackage{amsmath,amssymb,amsfonts}
\usepackage[colorlinks=false,hyperfootnotes=false]{hyperref}

\newcommand{\be}{\begin{equation}}
\newcommand{\ee}{\end{equation}}
\newcommand{\bel}[1]{\begin{equation}\label{#1}}
\newcommand{\ba}{\begin{eqnarray}}
\newcommand{\ea}{\end{eqnarray}}
\newcommand{\bal}[1]{\begin{eqnarray}\label{#1}}

\usepackage[usenames]{color}

\DeclareMathOperator\erf{erf}

\begin{document}


\title[Distinguishing compact binaries through GW signatures]{Distinguishing types of compact-object binaries using the gravitational-wave signatures of their mergers}

\author[I. Mandel and C.-J. Haster and M. Dominik and K. Belczynski]
{Ilya Mandel$^{1,2}$\thanks{E-mail: imandel@star.sr.bham.ac.uk} and Carl-Johan Haster$^{1,3}$ and Michal Dominik$^{4}$ and Krzysztof Belczynski$^{4}$\\
$^{1}$School of Physics and Astronomy, University of Birmingham, Birmingham, B15 2TT, United Kingdom\\
$^{2}$Monash Center for Astrophysics, Monash University, Clayton, VIC 3800, Australia\\
$^{3}$Center for Interdisciplinary Exploration and Research in Astrophysics (CIERA)\\ \& Dept. of Physics and Astronomy, 2145 Sheridan Rd, Evanston, IL 60208, USA\\
$^{4}$Astronomical Observatory, University of Warsaw, Al.~Ujazdowskie 4, 00-478 Warsaw, Poland}




\pagerange{\pageref{firstpage}--\pageref{lastpage}}\pubyear{2015}

\maketitle

\label{firstpage}


\begin{abstract}

We analyze the distinguishability of populations of coalescing binary neutron stars, neutron-star black-hole binaries, and binary black holes, whose gravitational-wave signatures are expected to be observed by the advanced network of ground-based interferometers LIGO and Virgo.  We consider population-synthesis predictions for plausible merging binary distributions in mass space, along with measurement accuracy estimates from the main gravitational-wave parameter-estimation pipeline.  We find that for our model compact-object binary mass distribution, we can always distinguish binary neutron stars and black-hole--neutron-star binaries, but not necessarily black-hole--neutron-star binaries and binary black holes; however, with a few tens of detections, we can accurately identify the three subpopulations and measure their respective rates.

\end{abstract}



\section{Introduction}

Merging binary neutron stars (NS-NS), binary black holes (BH-BH), and mixed black-hole -- neutron-star binaries (BH-NS) are among the most likely sources for the advanced gravitational-wave detectors LIGO \citep{AdvLIGO} and Virgo \citep{AdvVirgo}.  The rates could range from a detection every few years to a few hundred detections per year for each source type \citep{ratesdoc} once detectors reach full sensitivity toward the end of the decade, although the first detections are possible as early as the end of 2015 \citep{scenarios}.  One key question is whether it will be possible to determine the type of the binary --- NS-NS, BH-NS, or BH-BH --- from its gravitational-wave signature.

The surest way of distinguishing NS and BH binaries is via tidal dissipation in neutron stars leading to additional loss of orbital energy on top of that lost through gravitational waves (GWs), and their possible eventual tidal disruption.  However, the dominant GW signature through most of the LIGO-Virgo frequency band will be well-described by point-particle waveforms, with tidal effects becoming important only at higher frequencies where detectors are less sensitive, making it difficult to distinguish binary types in this way except, perhaps, for the loudest sources \citep[e.g.,][ and references therein]{Read:2013}.  Electromagnetic signatures in the form of short gamma-ray bursts, afterglows, and kilonovae could also help distinguish binaries with matter that can undergo tidal disruption (NS-NS or BH-NS) from those without (BH-BH); however, electromagnetic signatures may prove difficult to observe even for most NS-NS and BH-NS binary mergers \citep[e.g.,][]{Kelley:2012}, nor could they distinguish between those binary types given the current limited understanding of emission processes.  Other approaches to distinguishing subpopulations of different binary types in the overall set of GW observations would rely on classification based on clustering in the parameter space of all available information about the binaries, including spin, which appears to be observationally higher for BHs than NSs in compact binaries \citep[e.g.,][ and references therein]{MandelOShaughnessy:2010}.  However, in the absence of confident predictions for NS and BH spin, here we will focus on classification based exclusively on component mass.

Mass-based binary classification depends on (i) the assumptions about the mass population of NS and BH binaries, (ii) the accuracy with which component masses can be determined from the GW signature of an individual merger observation, and (iii) inference about the source population based on a statistical analysis of multiple detections.  In this Letter, we address all three of these issues in turn.  We argue that (i) astrophysical models predict that NS-NS, BH-NS, and BH-BH binary subpopulations may be strongly clustered, aiding source identification; (ii) a careful analysis of parameter measurement accuracy is required to take into account possible degeneracies; (iii) even if the binary type cannot be confidently determined from an individual measurement, a set of observations can be used to constrain mass distributions and measure the rates of different merger types.  We find that for our model of the compact-object binary mass distribution, based on population-synthesis simulations \citep{Dominik:2014}, we can always distinguish NS-NS and BH-NS binaries, but not necessarily BH-NS and BH-BH binaries.  However, with a few tens of detections, we can identify subpopulations and measure their respective rates to an accuracy dominated by the Poisson counting statistics of each binary type rather than contamination among subpopulations.

\section{Mass distributions}
\label{sec:binaries}
 
If the NS and BH mass distributions are directly adjacent, then it will be impossible to confidently determine the type of an individual source from finite-accuracy mass measurements alone.
However, observations of NS binaries, particularly via radio pulsars, and BH X-ray binaries indicate that there may be a mass gap between the two populations: the highest measured NS masses just exceed $2 M_\odot$ \citep{Lattimer:2012} while BH masses may only start at $\sim 4$ -- $5.5 M_\odot$ depending on the assumed shape of the distribution \citep{Ozel:2010,Farr:2010}; but see \citet{Kreidberg:2012} for possible selection biases that could push the BH masses lower.  This mass gap could be indicative of the supernova explosion mechanism \citep{Belczynski:2012}.

Previous analyses of the distinguishability of source populations based on GW observations typically relied on a mass gap individually applied to both components \citep[e.g.,][]{Hannam:2013}.  However, in practice, even if NSs occupy the mass space through 2 solar masses and BH masses start at 5 solar masses, it is not necessarily the case that merging compact binaries can contain arbitrary combinations of NS and BH masses within those respective ranges.  The compact-object mass distribution in binaries depends on the outcome of binary evolution.  To carry out a self-consistent analysis of the distinguishability of source populations, we consider sources that arise from an astrophysically modeled population of binaries and compare their mass measurements against the joint boundaries on component masses as given by the same population.

For this study, the binary population was generated with the StarTrack population synthesis code \citep{Belczynski:2008} under the ``Standard'' model B of \citet{Dominik:2012}.   This model builds on \citet{Belczynski:2008} by including the rapid supernova engine of \citet{Fryer:2012}, which allows for the formation of NSs with large natal kicks from single stars with zero-age main sequence (ZAMS) mass below $\sim 21 M_\odot$, and nearly complete fallback onto BHs with correspondingly low natal kicks from more massive progenitors, yielding minimum BH mass above $\sim 5 M_\odot$; realistic prescriptions for common-envelope binding energies and mass loss through stellar winds; potential binary disruptions in supernovae and stellar mergers within the common envelope; and a prescription for electron capture supernovae which lead $\sim 7$--$11 M_\odot$ ZAMS mass stars in binaries to form NSs with low natal kicks and masses near $1.25 M_\odot$; see \citep{Dominik:2012} for details.  We note that many of the input parameters and assumptions that go into population synthesis, such as those related to mass loss and mass transfer, BH supernovae fallback and kicks, and common envelope evolution, are poorly constrained.  Therefore, this is only one of many possible model variations; see the series \citep{Dominik:2012,Dominik:2013,Dominik:2014} for discussion, and \citep{PostnovYungelson:2014} for a recent review of decades of work in this field.  We do not insist on this model being accurate, but only use it for illustration purposes to show how general features of the model, and particularly clustering of the subpopulations in mass space, can be used to distinguish the compact binary type.  

\begin{figure}
\includegraphics[width=\columnwidth]{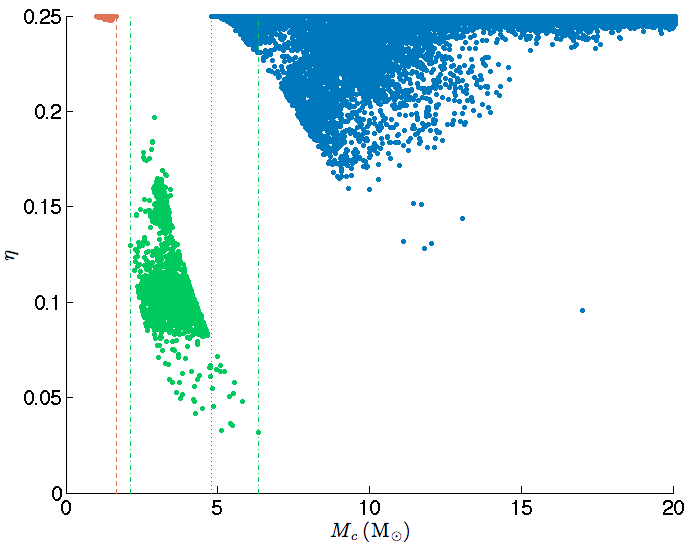}
\caption{\label{fig:binaries}
The mass distribution in $M_c-\eta$ space of compact-object binaries detectable by advanced gravitational-wave observatories as simulated by \protect\citep{Dominik:2014}, using the standard model of \protect\citep{Dominik:2012}.  NS-NS, BH-NS, and BH-BH populations occupy the top left (red), bottom middle (green) and top right (blue) regions of the plot, respectively. Each dot corresponds to a potentially detectable source, but the number of binaries plotted is increased far beyond the anticipated detection rate \protect\citep{ratesdoc} to better indicate the population distribution. Chirp mass boundaries for the three distributions are shown with dashed red, dash-dotted green, and dotted blue vertical lines, respectively.  BH-BHs with chirp mass above $20 M_\odot$, which cannot be confused for another source type, are omitted from the figure.}
\end{figure}

Figure \ref{fig:binaries} shows the distribution of masses for the three binary types among the binaries detectable by the advanced-detector network as estimated by \citep{Dominik:2014}.  A wide range of metallicities is included in these calculations to simulate the redshift-dependent metallicity distribution in the Universe; in fact, as \citep{Dominik:2014} showed, the local BH-BH mergers are dominated by low-metallicity massive binaries that formed in the early Universe, with long time delays between star formation and merger.  

We re-parametrize the component masses $m_1$, $m_2$ via the chirp mass $M_c=m_1^{3/5} m_2^{3/5} (m_1+m_2)^{-1/5}$, which is generally well-measured with gravitational waves, and the symmetric mass ratio $\eta=m_1 m_2  (m_1+m_2)^{-2}$.   The main features of the model become apparent from the figure: the three source types are in general clearly separated in the chirp mass --- mass ratio space.  

NS-NS systems are very strongly clustered near equal masses ($\eta > 0.2477$, or $m_2/m_1 > 0.82$).  This clustering of the NS-NS mass ratio reflects their nearly equal-- (and typically low--) mass progenitor binaries, preferred since they allow the primary to form through an electron-capture supernova which avoids disrupting the binary; mass transfer after the first supernova leads to a growth in the mass of the primary by no more than $\sim 7\%$, keeping the mass ratio nearly equal.    BH-BH systems also cluster near equal masses at solar metallicity, in part because a large difference in initial masses causes unstable mass transfer and a common-envelope merger during the first mass transfer phase, and in part because of the very efficient mass loss through stellar winds which tends to cluster BH masses despite a very broad range of ZAMS masses.  Mass loss rates are reduced for lower metallicities; therefore, more unequal mass ratios are possible for lower metallicities, with mass ratios down to $\eta \sim 0.1$ for BH-BHs that formed at $0.01 Z_\odot$.    Meanwhile, BH-NS systems have more extreme mass ratios, ranging between $\eta \sim 0.03$ and $\eta \sim 0.2$, consistent with the supernova fallback prescription that naturally generates the mass gap.

The chirp mass is lowest for NS-NS systems given the constraints on the maximum component masses.  In fact, there's a gap between the highest NS-NS chirp mass ($M_c=1.67 M_\odot$, for a $1.92 M_\odot + 1.91 M_\odot$ binary formed at $0.1 Z_\odot$) and the lowest BH-NS chirp mass ($M_c=2.11 M_\odot$, for a $6.10 M_\odot + 1.10 M_\odot$ binary formed at $0.75 Z_\odot$).  Meanwhile, BH-BH systems do overlap with BH-NS binaries in chirp mass, as the lowest-mass equal-mass BH-BH systems ($5.55 M_\odot + 5.53 M_\odot$ forming at $0.025 Z_\odot$) can have a lower chirp mass ($M_c = 4.82 M_\odot$) than the highest-chirp mass, extreme-mass-ratio BH-NS system ($M_c = 6.42 M_\odot$ for a $48.0 M_\odot + 1.70 M_\odot$ binary that formed at  $0.005 Z_\odot$).  

\section{Measurement accuracy}

We estimate the accuracy of mass measurements on individual events through the \texttt{LALInference} parameter-estimation toolkit, designed for efficient stochastic exploration of the parameter space of GW signals \citep{Veitch:2014}.  We eschew the large-scale mock data challenges attempted elsewhere \citep[e.g.,][]{Singer:2014,Berry:2015} and instead consider several events placed at the boundaries of the regions described in the previous section.  

We analyze mock injections of the highest-chirp-mass NS-NS system labeled as detectable in our population-synthesis simulation, with ($1.92 M_\odot $, $1.91 M_\odot$) components; the lowest-chirp-mass BH-NS system with ($6.10 M_\odot $,  $1.10 M_\odot$) components; the highest-mass-ratio BH-NS system with ($5.60 M_\odot$, $2.10 M_\odot$) components; the highest-chirp-mass BH-NS system with ($48.0 M_\odot$,  $1.70 M_\odot$) components; and the lowest-chrip-mass BH-BH system with ($5.55 M_\odot$, $5.53 M_\odot$) components.  Given the uncertainty about NS and BH spins, we consider four variations of each injection: non-spinning components, components with aligned spins, and spinning components with two randomly chosen arbitrary spin directions.  In all cases when components are spinning, the BH dimensionless spin parameter is $\chi=0.8$; the NS is not spinning in BH-NS injections and only the higher-mass NS is spinning with a dimensionless spin parameter of $\chi=0.05$ in spinning NS-NS injections (representative of the fastest known NS spin in a confirmed merging Galactic NS-NS binary, J0737-3039A).  For NS-NS and BH-BH systems, as well as the highest-mass-ratio BH-NS system, we use for all injections and templates the IMRPhenomP precessing waveform family \citep{Hannam:2013waveform}, which models all phases of the waveform: inspiral, merger, and ringdown.  However, since this waveform model is restricted to mass ratios less extreme than $1:4$, i.e., $\eta\ge 0.16$, BH-NS injections (other than the high-mass-ratio injection) are made and analyzed with SpinTaylorT4 waveforms that include only the post-Newtonian inspiral phase \citep{BuonannoChenVallisneri:2003b}, not the merger and ringdown \citep[see][ for an analysis of BH-BH signals with SpinTaylor waveforms, which shows somewhat greater typical mass measurement uncertainty]{Littenberg:2015}. 

In all analyses, we allow for arbitrary, precessing spins in the templates.  We use flat priors on the component masses, flat priors on dimensionless spin magnitudes $\chi \in [0,1]$ ($\chi \in [0,0.9]$ for IMRPhenomP analyses because of limitations in that waveform's region of validity)
and isotropic priors on spin directions for the Bayesian analysis.  All analyses are carried out from a starting frequency of $40$ Hz with mock signal-only data from a LIGO--Virgo detector network operating at advanced LIGO sensitivity in the zero-detuned, high-power configuration \citep{PSD:AL}.  The injected distance was chosen to yield a network signal-to-noise ratio of 12, roughly at the lower limit of detectability \citep[e.g.,][]{scenarios,Berry:2015}, thereby yielding conservative predictions on parameter-estimation accuracy.
 
\begin{figure}
\includegraphics[width=\columnwidth]{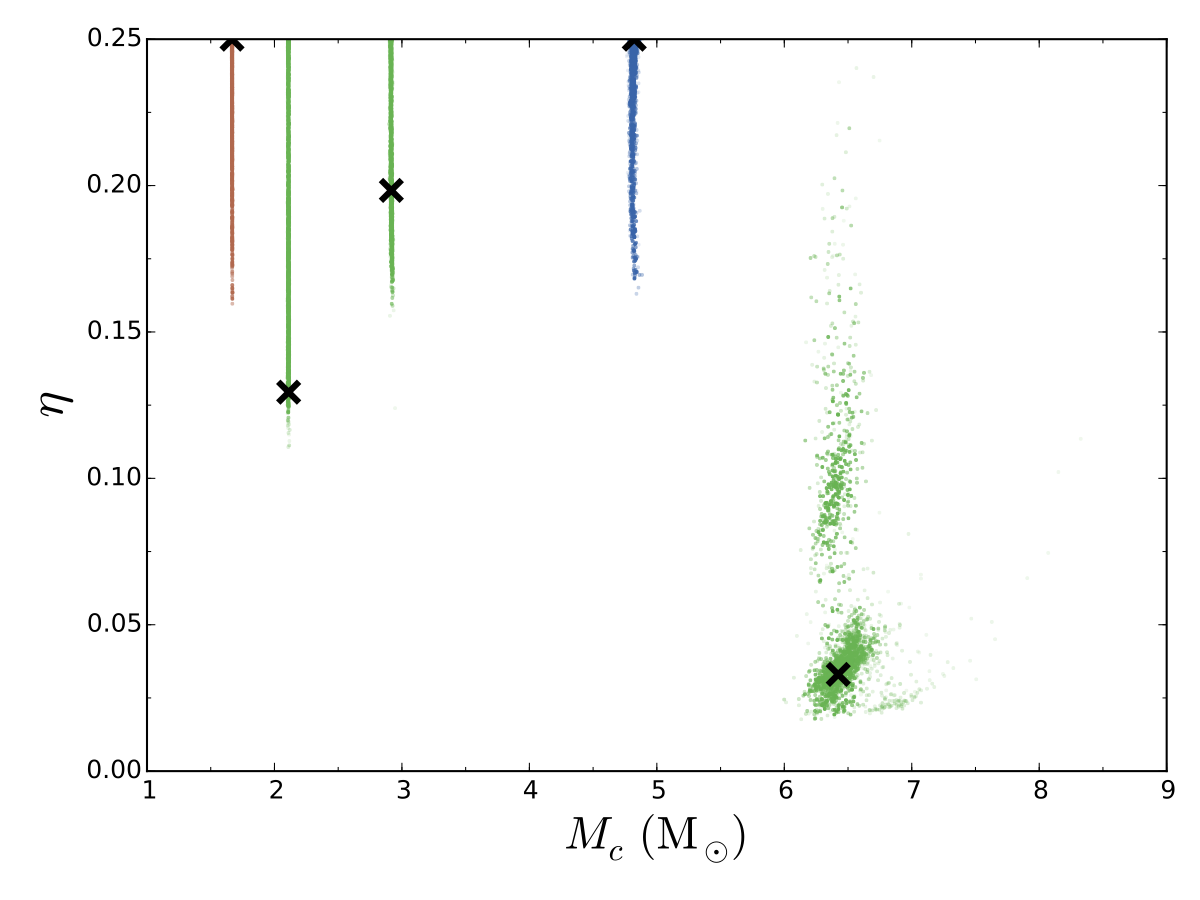}
\caption{\label{fig:PEall}
Posterior PDF samples for the 20 injections described in the text: NS-NS (red), BH-BH (blue), and three BH-NS (green), each with four spin variants.  Point opacity corresponds to the relative value of the posterior probability density function.  Chirp mass measurements are typically very accurate (with the exception of the very massive, extreme-mass-ratio BH-NS binary on the bottom right of the plot), while mass ratios are partly degenerate with spins, and their measurement accuracy depends on the exact injected spin configuration.}
\end{figure} 

Figure \ref{fig:PEall} shows the posterior probability density functions (PDF) for the 20 analyses described above (four injections -- non-spinning, spin-aligned, and two precessing -- for each of 5 events).  Several thousand samples for each event are drawn from the posterior by \texttt{LALInference}; the opacity of plotted points rises with the value of the posterior PDF.  The injected masses are denoted with black crosses. 

The chirp masses are typically very well measured (posteriors look like narrow nearly-vertical lines), except for the extreme-mass-ratio, high-mass BH-NS injection that appears on the bottom right of the plot.  This is consistent with other analyses \citep[e.g.,][]{S6PE,Rodriguez:2013BNS,Hannam:2013}, and is readily understood from the governing role played by the chirp mass in the phase evolution of the binary during the inspiral \citep[e.g.,][]{PNwaveforms:2009,Ohme:2013}.  The 90\% credible regions in chirp mass spanning all four spin variations for the given component mass combinations are $\lesssim 0.01 M_\odot$ for the NS-NS injections and $\lesssim 0.1 M_\odot$ for the BH-BH injections.  For BH-NS injections, the 90\% credible regions in chirp mass span $\lesssim 0.03 M_\odot$, except for the very poorly measured extreme-mass-ratio, high-mass injections discussed below, for which the combined 90\% credible region in chirp mass spans $\sim 2 M_\odot$.  Hence, measurement of the chirp mass alone allows us to distinguish NS-NS and BH-NS binaries, which are separated by a chirp-mass gap of $0.45 M_\odot$ -- much larger than the chirp-mass measurement uncertainty -- if we trust the model described in the previous section.  Conversely, a single source falling into the chirp-mass gap between NS-NS and BH-NS binaries would allow us to disprove this model.  Of course, we do not necessarily trust the population-synthesis predictions for the exact boundaries of the NS-NS, BH-NS, and BH-BH clusters in the mass plane; this will be addressed in the following section.

\begin{figure}
\includegraphics[width=\columnwidth]{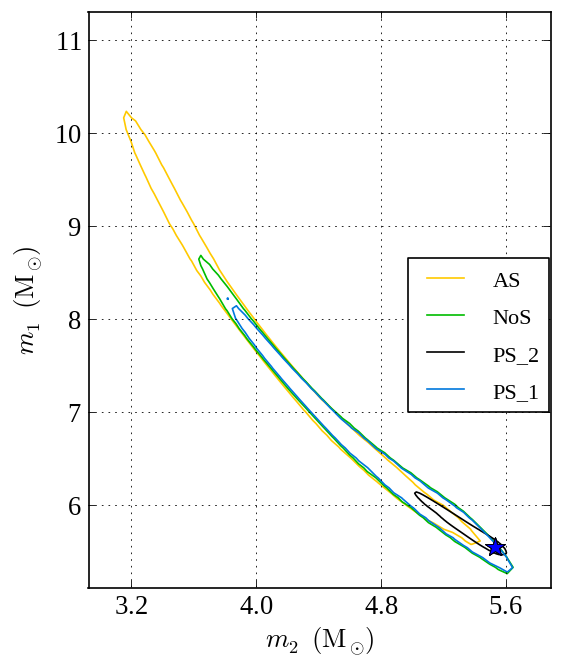}
\caption{\label{fig:BH-BH}
90\% credible regions for the four BH-BH injections in component mass space ($m_1$ defined to be the larger component mass, $m_1 > m_2$).  All injections have $5.55 M_\odot$, $5.53 M_\odot$ components, but differ in spin magnitudes (0 for NoS, 0.8 for each of the dimensionless spins for the other injections) and directions (aligned for AS, random and precessing for the two PS injections).  Mass measurements are more accurate when precession breaks degeneracies between intrinsic parameters.  The second precessing injection may be atypically favorable for mass measurement 
\citep[cf.][]{Littenberg:2015} and is shown as an example of what could be achieved for a nearly edge-on precessing binary.
}
\end{figure} 

On the other hand, the mass ratio $\eta$ is more poorly measured, and partly degenerate with spin \citep[e.g.,][]{Ohme:2013}.  The difficulty of measuring the mass ratio is apparent from the very elongated posteriors in Fig.~\ref{fig:PEall}, and from the typical ``banana-shaped'' plots of posteriors in component-mass space shown in Fig.~\ref{fig:BH-BH}.  The latter plot moreover illustrates the dependence of measurement accuracy on the configuration of spins in the injected system.  Systems with precession typically have the best-measured masses at a fixed signal-to-noise ratio, as evidence of precession allows for more accurate spin measurements, partly breaking the mass-ratio--spin degeneracy \citep{vanderSluys:2008b,Chatz:2014,Chatz:2015}.  The limited accuracy of measuring the mass ratio $\eta$ means that posteriors for BH-BH and BH-NS events could overlap, making it difficult to distinguish between the two.  However, our present attempts to estimate the extent of the posterior for massive, extreme-mass-ratio BH-NS binaries that could overlap with the BH-BH distribution are limited by the lack of waveforms that include both precession and mergers and ringdowns for such systems.  By injecting and analyzing the BH-NS system on the bottom right of Fig.~\ref{fig:PEall}, which has a total mass of $\sim 50 M_\odot$, with inspiral-only waveforms, we are losing significant information and may be artificially extending the measurement uncertainty, creating, in our plot color scheme, ``the green-eyed monster which doth mock''.  Moreover, if we are sufficiently fortunate to have significant precession in the binaries \citet[see Fig.~\ref{fig:BH-BH} and][]{Chatz:2015}, the posteriors will be narrowed and more accurate parameter estimation will be possible.  Furthermore, we considered parameter estimation at a signal-to-noise ratio near the threshold for detectability \citep[e.g.,][]{scenarios, Berry:2015}; louder signals will lead to more accurate inference. 

In any case, if the population boundaries are known in advance (e.g., if we have sufficient trust in population synthesis models models of Sec.~\ref{sec:binaries}) we anticipate being able to readily distinguish $\gg 99\%$ of the BH-NS and BH-BH populations on the basis of chirp mass measurements alone for the model we have considered: the 99th percentile of the BH-NS chirp-mass distribution, $4.76 M_\odot$, is readily distinguishable from the 1st percentile of the BH-BH chirp-mass distribution, $5.67 M_\odot$, given chirp mass measurement uncertainties $\lesssim 0.1 M_\odot$.

\section{Population statistics}

We now focus on the question of distinguishing types of compact binary sources within the observed population.  Given the uncertainty in the claimed range of a putative mass gap between NS and BH masses, and the fact that the mass combinations of merging binaries may be more restrictive than suggested by the mass gap, as discussed in Sec.~\ref{sec:binaries}, we address the question of {\it classification using only the observed population itself}, rather than relying on any a priori assumptions about subpopulation boundaries or mass gaps.  We provide approximate scaling arguments and typical useful constraints.  The exact accuracy of classification depends on the details of the true distribution; the plausible two-dimensional mass distribution described in Sec.~\ref{sec:binaries} is used as an illustration.

{\it Measuring a distribution boundary from the observed population.}  A boundary of a distribution, if abrupt, is generally easier to measure than other distribution parameters \citep[cf.][]{Mandel:2014}.  For example, with perfect accuracy on individual measurements, the accuracy of the estimate of the mean of a sampled distribution scales inversely with the square root of the number of observations (samples) -- but the location of a sharp boundary in the distribution can be estimated with an accuracy that scales inversely with the number of observations.  If observations have an individual measurement uncertainty $\sigma$, and $N$ observations are spread over a parameter range $\Delta$, an abrupt boundary can be estimated with accuracy $\sim \sqrt{\frac{\sigma^2}{N} + \frac{\Delta^2}{N^2}}$.  Of course, the boundary is harder to measure if the edge is gradual rather than abrupt.  However, our specific two-dimensional mass distributions do, in fact, have fairly sharp one-dimensional boundaries following suitable coordinate transformations; e.g., in our model, the mass of the lower-mass BH-BH companion has a boundary of $m_2 \gtrsim 5.5 M_\odot$.  The smaller companion mass has a one--sigma uncertainty of $\sim 1.2 M_\odot$ for comparable-mass binaries (taking the worst-case credible region of  Fig.~\ref{fig:BH-BH}, which shows 90\% credible intervals),
and the BH-BH $m_2$ population width is $\Delta \sim 10 M_\odot$, so $\sim 10$ observations confidently known to be BH-BHs are needed to constrain the lower boundary on BH mass in the BH-BH distribution to an accuracy of $\lesssim 1 M_\odot$.  


{\it Establishing the existence of a break in the distribution.}  Observing the existence of a break between subpopulations need not rely on very precise individual measurements.  In fact, even perfectly accurate measurements cannot individually confirm the presence of break in the distribution (although a single perfect measurement that falls into an expected break would disprove its existence).  Rather, the existence of a break is indicated by a drop in the density of observations in the parameter range of the break.  Critically, this is observable even when the parameter measurement accuracy is comparable to or even exceeds the width of the break.  Denoting the width of the break by $\Delta_\textrm{break}$ (as opposed to the net distribution width $\Delta$), the measurement error in the parameter direction across the break by $\sigma$, and the total number of observations by $N$, the expected number of observations in the break is $\sim N \frac{\sigma}{\Delta}  \erf \left({\frac{\Delta_\textrm{break}}{\sqrt{2} \sigma}}\right)$.  By comparison, if there were no break in the underlying distribution, $N \frac{\Delta_\textrm{break}}{\Delta}$ observations would be expected in that region.  Therefore, the dearth in the number of observations in the break will be significant when the difference between the actual number of observations and $N \frac{\Delta_\textrm {break}}{\Delta}$ is large relative to the Poisson fluctuations in the latter, $\sqrt{N \frac{\Delta_\textrm{break}}{\Delta}}$.
[This does not indicate an absolute break, which cannot be established with a finite number of observations, but a significant local drop in the probability density function of the inferred underlying distribution.]  Again, using the break between BH-NS and BH-BH distributions as an example, with $\Delta_\textrm{break} \approx 3.3 M_\odot$ in $m_2$, approximately 60 observations are required for a confident detection of a break at a three-sigma significance.  In a contemporaneous study, \citet{Littenberg:2015} find that hundreds of observations may be necessary to measure a break or gap in the mass distribution when considering an ad hoc population that is flat in component masses; this is consistent with the application of our analysis to their distribution. 

{\it Classifying observations into sub-populations.}  As mentioned earlier, we do not wish to rely on a priori divisions of the compact-binary mass parameter space into regions, but instead to measure these regions from observations.  \citet{Mandel:2010stat} described the statistical procedure for inference on population parameters based on a limited set of uncertain observations.  \citet{Farr:2013} introduced a classification process based on models for subpopulation distributions whereby model parameters are fit as part of the classification process.  Meanwhile, a number of unmodeled or weakly modeled schemes for extracting a distribution and classifying subpopulations exist, ranging from Dirichlet processes to k-means clustering (using the weighted posterior probability density functions of individual observations).  The preferred algorithm depends on the priors that one wishes to place on the population distribution: modeled approaches require a choice of the shape (model) of the underlying distribution, which can aid measurement accuracy over unmodeled classification schemes at the expense of introducing bias if the model shape does not match the observed population. In general, however, clustering should be successful whenever the presence of a break in the distribution can be established.  As discussed above, for the BH-NS -- BH-BH division this requires several tens of observations of each binary type if the distribution is similar to that described in Sec.~\ref{sec:binaries}.  By contrast, the NS-NS and BH-NS subpopulation are cleanly separated in chirp mass, i.e., the measurement uncertainty $\sigma$ is very small relative to the size of the break in the chirp-mass distribution.  In this case, the existence of a break can be established as long as $\sqrt{N \frac{\Delta_\textrm{break}}{\Delta}}$ is at least a few, i.e., $\sim 10$ observations are sufficient.  

{\it Counting sources.}  Even if the subpopulations are classified and their boundaries are accurately determined, individual sources cannot necessarily be perfectly classified as belonging to a given subpopulation.  Nonetheless, the number of sources in each subpopulation can be measured with some accuracy \citep{Farr:2013}.   The total mis-classified source fraction due to the uncertainty in cluster boundaries is  $\lesssim \sqrt{\frac{\sigma^2}{N} + \frac{\Delta^2}{N^2}} \Delta^{-1}$ --- which, for the typical parameters we have considered here, is less than the fractional Poisson fluctuation in the subpopulation counts, $\sim \frac{1}{\sqrt{N}}$.  Therefore, we expect estimates of merger rates for various binary types to be limited by Poisson counting statistics on the number of events rather than classification errors.

\section{Summary and future directions}

We considered a plausible distribution of masses of merging NS-NS, BH-NS, and BH-BH binaries, whose gravitational-wave signatures would be detectable by advanced ground-based gravitational-wave interferometers LIGO and Virgo.  We found that NS-NS and BH-NS binaries were clearly separated in chirp mass, while BH-NS and BH-BH distributions had some overlap in chirp mass, but were widely separated in the mass ratio.  We evaluated the measurement accuracy of mass parameters using the \texttt{LALInference} toolkit that will be used for parameter estimation on LIGO-Virgo candidates, considering a variety of spin configurations.  We concluded that if the mass distributions are known in advance, there will be no confusion on the NS-NS -- BH-NS interface, and less than 1\% of binaries could be incorrectly classified on the BH-NS -- BH-BH interface.  We then considered the case where mass distributions are not known in advance, and determined that a few tens of detections are sufficient to correctly cluster the subpopulations in the two-dimensional mass space if the underlying population is similar to the standard model of \citet{Dominik:2014}.  Although individual sources may still be mis-classified, we anticipate that this will not be a significant source of error in the estimates of merger rates for various subpopulations.

We considered a population synthesis model in which the supernova prescription, particularly the amount of fallback on a newly formed BH, is designed to reproduce the apparent observational mass gap between NS and BH masses.  If the NS and BH mass distributions are instead assumed to be continuous, as in the delayed supernova model of \citet{Belczynski:2012}, the quantitative results change:  NS-NS and BH-NS chirp mass distributions overlap \citep{Dominik:2014}.  However, the first percentile of the BH-NS chirp mass distribution is still significantly larger than the 99th percentile of the NS-NS chirp mass distribution, and similarly for the second and 98th percentiles of the BH-BH and BH-NS chirp mass distributions, respectively.  In fact, the subpopulations are sufficiently concentrated away from the boundaries that the conclusions we have reached remain robust, and clustering for the delayed supernova model is still possible with tens of observations.

Another complication could be the low number of observations of a particular species of binaries; for example, in the high BH supernova kick model of \citet{Dominik:2014}, rates of BH-NS binaries are very strongly suppressed relative to the other binary types.  In this case, the number of BH-NS detections may be too small to identify a distinct cluster and appropriately classify them in the absence of a priori distribution models.

This Letter shows that, at least for the physical models considered, it should be possible to cluster and classify populations of NS-NS, BH-NS, and BH-BH merging binaries based on their gravitational-wave signatures.  However, further rapid progress is required to address a number of practical challenges given the short timescale to anticipated GW detections \citep{ratesdoc,scenarios}.  Actual clustering and classification schemes need to be implemented and tested for robustness on a variety of plausible source populations, evaluating the relative trade-offs of modeled approaches which require a choice of the shape (model) of the underlying distribution \citep{Farr:2013} vs.~unmodeled classification schemes which are less sensitive to a priori assumptions about the distribution shape.  Subpopulation rate estimates must take selection biases into account \citep{Berry:2015,Mandel:2015selection}.  The impact of other sources of parameter-estimation errors \citep[e.g., systematics due to imperfect waveform knowledge,][]{S6PE,Favata:2014} needs to be evaluated.  Classification should also be extended to consider other parameters beyond component masses, such as spins and spin-orbit misalignment angles \citep[e.g..][]{Vitale:2014}, as well as additional subpopulations, such as dynamically formed binaries in dense stellar environments.  

\section*{Acknowledgments\\}

IM acknowledges STFC funding and the hospitality of the Monash Center for Astrophysics supported by a Monash Research Acceleration Grant (PI Y.~Levin).  CJH acknowledges support from CIERA and an RAS grant.   MD and KB acknowledge support from the Polish Science Foundation ``Master2013'' Subsidy, by Polish NCN grant SONATA BIS 2.  MD acknowledges support from the National Science Center grant DEC-2011/01/N/ST9/00383.
We are grateful for computational resources provided by the Leonard E Parker Center for Gravitation, Cosmology and Astrophysics at University of Wisconsin-Milwaukee. 
MD and KB would like to thank the N.~Copernicus Astronomical Centre in Warsaw, Poland, and
the University Of Texas, Brownsville, TX, for their courtesy, enabling us to use their 
computational resources.   IM and KB are grateful to the Aspen Center for Physics, supported by NSF grant \# 1066293, where some of the discussions leading to this Letter took place.    We are grateful to Tyson Littenberg, Ben Farr, Vicky Kalogera, Daniel Holz, Will Farr, and Alex Nielsen for comments and suggestions.

\bibliographystyle{hapj}
\bibliography{Mandel}

\label{lastpage}
\end{document}